\documentclass{svproc}
\usepackage{url}
\usepackage{graphicx}

\usepackage{amssymb}

\begin{document}
\mainmatter              
\title{Theory Summary at Strangeness in Quark Matter 2019}
\titlerunning{TheorySQM19}  
%
\author{Jacquelyn Noronha-Hostler\inst{1} }
\authorrunning{Noronha-Hostler} 
%
\tocauthor{Jacquelyn Noronha-Hostler}
\institute{University of Illinois at Urbana-Champaign, Urbana, IL 61801, United States,\\
\email{jnorhoss@illinois.edu}}

\maketitle              

\begin{abstract}
This is the theory summary of Strangeness in Quark Matter 2019 conference. Results include the state-of-the-art updates to the Quantum Chromodynamics (QCD) phase diagram with contributions both from heavy-ion collisions and nuclear astrophysics, studies on the QCD freeze-out lines, and several aspects regarding small systems including collectivity, heavy flavor dynamics, strangeness, and hard probes.
\end{abstract}
\section{Introduction}


One of the crucial signatures for the discovery of the Quark Gluon Plasma (QGP) was the measurement of strangeness enhancement due to the ease of producing more strangeness particles from gluon interactions or an annihilation of a light quark anti-quark pair. Since that time the study of strangeness has evolved significantly. For instance, now that it is understood that the Quantum Chromodynamic (QCD) phase transition is a cross-over \cite{Aoki:2006we}, contrasting observables of light versus strange hadrons can provide insight into properties of this transition. Connected to the same QCD phase diagram but in the baryon rich regime, the interactions of strange hadrons are a necessary input to the QCD equation of state and can put constraints on the mass radius relationship of neutron stars. 
 
The natural next step is then studying the charm quark, which may not be thermalized with the rest of the QGP and can provide orthogonal information about its properties. In fact, charm quarks appear to be a particularly interesting probe in small systems \cite{Sirunyan:2018toe} and may provide key information to determine the limits of size of the QGP. However, one should caution that it is important to use realistic medium in theoretical descriptions, otherwise the results may be misleading. 

In this theory summary of Strangeness in Quark Matter 2019, I provide an overview of the latest breakthroughs in relativistic heavy-ion collisions and nuclear astrophysics that give insight into strange and charm quarks.

\section{QCD Phase Diagram: from Heavy-Ion Collisions to Neutron stars}

\begin{figure}
	\centering
	\begin{tabular}{c c}
\includegraphics[width=0.45\linewidth]{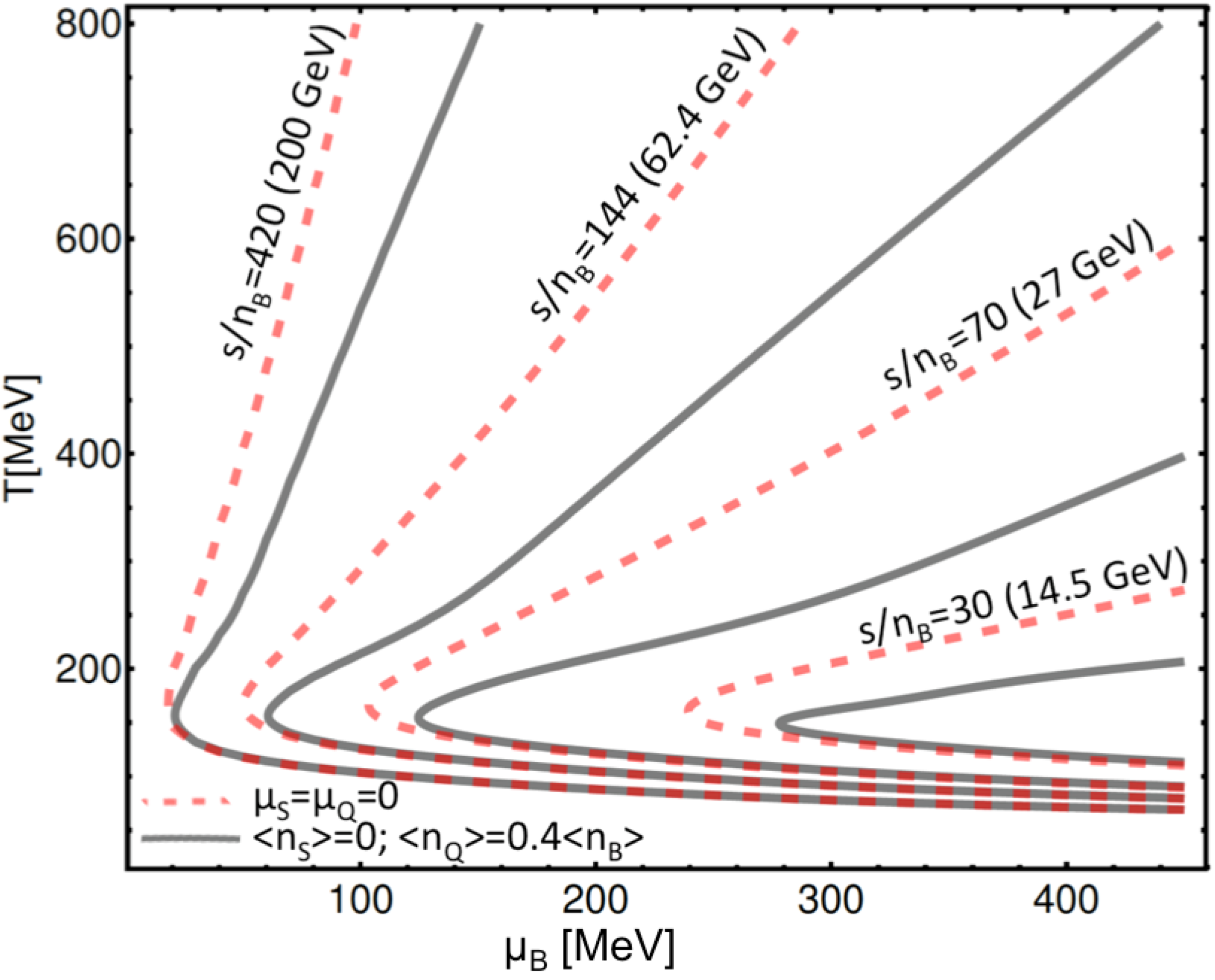} & \includegraphics[width=0.55\linewidth]{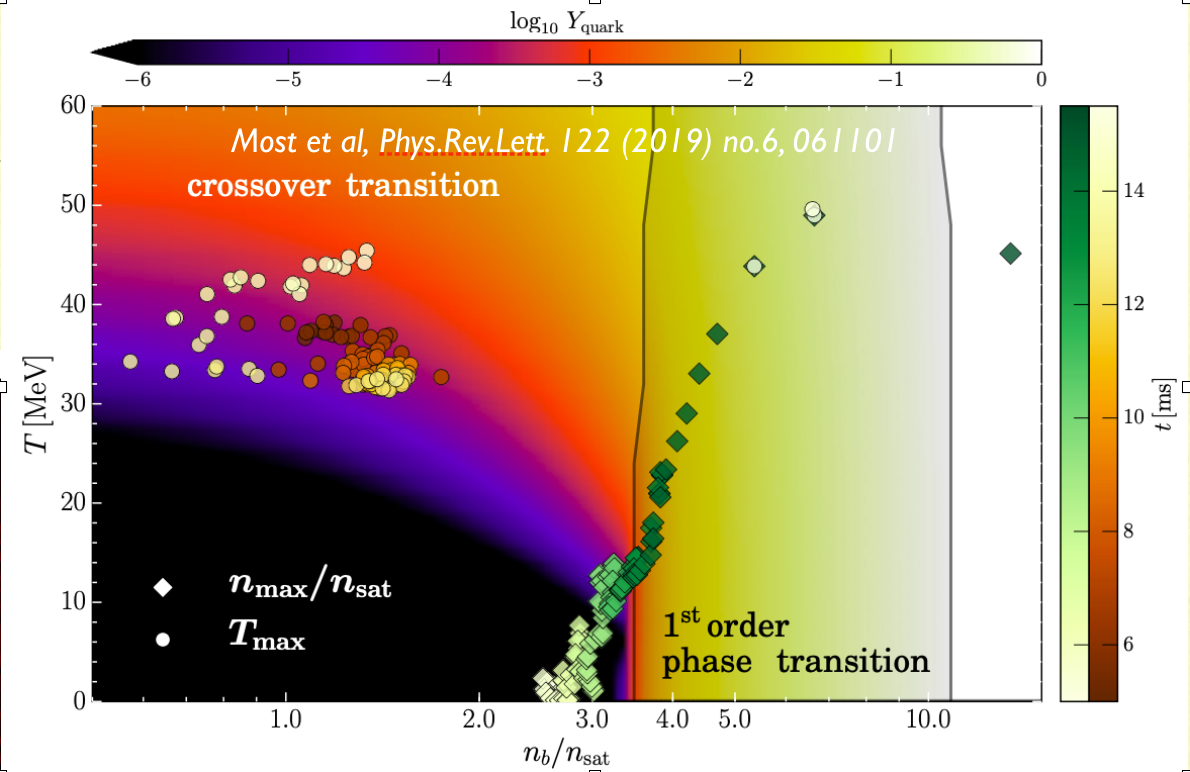} 
	\end{tabular}
	\caption{QCD phase diagram from Lattice QCD (left) with BSQ conserved charges from \cite{Noronha-Hostler:2019ayj}.   On the right is the estimated range in the QCD that neutron star mergers may reach using ideal relativistic hydrodynamic calculations coupled to GR \cite{Most:2018eaw}.}\label{fig:phases}
\end{figure}

Over the last few years significant advancements have been made to the QCD phase diagram.  The left side of Fig.\ \ref{fig:phases} demonstrates the most recent reconstruction of the QCD equation of state based on Lattice QCD results for an equation of state with three conserved charges: baryon number, strangeness, and electric charge where one finds that the assumption of strangeness neutrality pushes experiments to larger baryon chemical potentials \cite{Noronha-Hostler:2019ayj}. This implies that relativistic heavy ion collisions may possibly be closer to nuclear astrophysics than originally expected.  In nuclear astrophysics  it is possible to reach quite high temperatures in neutron star mergers (see Fig.\ \ref{fig:phases} on the right) and gravitational waves may provide hints if  deconfined matter is at the core of neutron stars \cite{Most:2018eaw}.

At $\mu_B=0$ the phase transition is a cross-over \cite{Aoki:2006we} and it is anticipated that at larger baryon densities a critical point may be discovered followed by a first-order phase transition line that may be reached in neutron star mergers or proto-neutron stars.  One of the primary signals of such a critical point is the kurtosis of net-proton fluctuations \cite{Stephanov:2011pb,Critelli:2017oub}.  However, caveats exist once one considers finite size effects, centrality binning, and detector efficiencies \cite{Sombun:2017bxi,Nouhou:2019nhe}. 

One of the crucial questions in nuclear astrophysics is: what is the state of matter at the core of a neutron star? Is it deconfined matter, just protons/neutrons, or strange baryons (and their non-trivial interactions)?   A possible signal for deconfiment would be the measurement of mass twins, which are stars that have the same mass but vastly different radii \cite{Alford:2013aca,Benic:2014jia,Montana:2018bkb}. Additionally, it  is important to properly understand repulsive versus attractive hyperon interactions since the addition of hyperons can affect the mass radius relationship \cite{Chatterjee:2015pua,Vidana:2018bdi,Ribes:2019kno}.  `

Unlike in most studies of phase transitions in fields like condensed matter, in nuclear physics the system may be far from equilibrium and transport coefficients play a significant role in the search for a critical point/first order phase transition. Shear and bulk viscosity need to be calculated at finite baryon densities \cite{Moreau:2019vhw}.  In heavy-ions where BSQ conserved charges are relevant, this also then leads to three diffusion transport coefficients for each conserved charges, which have been thus far been calculated in kinetic theory or non-conformal holographic models \cite{Bhadury:2019xdf,Rougemont:2015ona,Rougemont:2017tlu,Greif:2017byw,Denicol:2018wdp,Martinez:2019bsn}. Unlike in heavy-ions, in neutron star mergers the transport coefficients stem from weak interactions \cite{Alford:2019kdw,Alford:2017rxf}, which must eventually be incorporated into relativistic hydrodynamic calculations coupled to general relativity \cite{Bemfica:2019cop} (specifically bulk viscosity).

\subsection{Freeze-out}

While each event in heavy-ion collisions (or  single neutron star merger\footnote{I am currently unaware of an equivalent measurement to freeze-out in neutron star mergers but hypothesize that nuclei abundances would be a potential candidate.}) passes through the phase diagram in a unique manner depending on its initial conditions (expanding and cooling over time), one can measure the point of chemical freeze-out using identified particle yields \cite{Bellini:2018khg,BelliniSQM19} and fluctuations \cite{Critelli:2017oub,Borsanyi:2014ewa,Alba:2014eba,Noronha-Hostler:2016rpd}. 

There is a tension between yields of light and strange particles in hadronic yield comparisons with thermal fits \cite{BelliniSQM19}.  Fluctuations of conserved charges  \cite{Bellwied:2018tkc,Bluhm:2018aei} demonstrate a preference for a flavor hierarchy i.e. strange hadrons freezing out at a higher temperature than light hadrons. Additionally, transport models reveal a similar conclusion \cite{Bugaev:2018klr}. To explore this further, STAR has measured cross-correlations between conserved charges using certain identified hadrons \cite{Adam:2019xmk}.  However, it may be that there are alternative proxies for BSQ conserved charges that would be a better for direct comparisons to Lattice QCD \cite{Bellwied:2019pxh}. 

Following chemical freeze-out, one expects that kinetic freeze-out is achieved (although at very low beam energies the two appear to converge to the same temperature \cite{Adamczyk:2017iwn}). Recently, HADES used virtual photons to estimate temperatures reached at very low beam energies \cite{Adamczewski-Musch:2019byl} and found that the temperatures may be somewhat higher than originally expected.  

One challenge to the thermalization picture comes from the recently measured light nuclei that  appear to follow yields calculated from thermal fits.  Questions remain in terms of interpreting these results and their implications \cite{Oliinychenko:2018ugs,Cai:2019jtk}. 

\subsection{Hadron spectrum}

Understanding both the total number of possible hadrons and their interactions are fundamental to both the understanding of the hadron gas phase in heavy ion collisions and the composition of the core of neutron stars.  Recently, partial pressures were used to constrain the particle spectrum \cite{alba:2017mqu} and it was found that even resonances with the most experimental uncertainty are needed to reproduce Lattice QCD results.  However, the inclusion of in-medium effects of the HRG could also reproduce these partial pressures \cite{Aarts:2018glk}.  Additionally, significant progress has been made in understanding hyperon interactions in Lattice QCD \cite{Hiyama:2019kpw}, which may have wide reaching effects.

\section{Heavy Flavor and Hard Probes}

Because of the large mass of the heavy quarks, charm quarks should likely have much longer thermalization times \cite{Moore:2004tg}.  To understand the degree of thermalization of the charm quarks,  the effect of charm conservation was studied \cite{Zhao:2019fth}  and predicts a very large differences in the $D_s/D_0$ ratio, if charm is conserved.

Since the very first event-by-event heavy flavor \cite{Nahrgang:2014vza} and hard probe calculations \cite{Noronha-Hostler:2016eow}  many new observables have been proposed that correlate the soft and hard/heavy sectors \cite{Betz:2016ayq,Prado:2016szr}. For instance, it was first suggested in \cite{Jia:2012ez} that the event plane angle of higher order harmonics or hard probes would be less and less correlated with the soft event plane angle.  This was further confirmed in \cite{Betz:2016ayq,Prado:2016szr,Katz:2019fkc}.  Then in \cite{Plumari:2019yhg} a new correlation function was proposed in order to study the interplay with the soft and heavy flavor sectors, which also found that the higher harmonics are less correlated with the soft sector. 

One important caveat in most heavy flavor studies is that the hydrodynamic background can play a significant role if not tuned properly to the soft sector.  In \cite{Cao:2018ews} it was found that multiple heavy flavor models may appear to all simultaneously match both $R_{AA}$ and $v_2$, however, once identical backgrounds were taken for all models wide variations were seen comparing to the same observables. In \cite{Katz:2019fkc} two different choices in initial conditions were compared to experimental data and $v_2\{4\}/v_2\{2\}(p_T)$ of D mesons appeared to be the best choice to distinguish between the two initial condition models. 

Heavy flavor studies are sensitive to coalescence and fragmentation  \cite{Minissale:2019gbf}.  Additionally, understanding the origin of the heavy flavor transport coefficients in a strongly coupled \cite{Hambrock:2018olg} versus weakly coupled approach is important to understanding the properties of the QGP. 
While the soft gluon approximation is well motivated, it was found that its effect on suppression is negligible \cite{Blagojevic:2019qum}.  Finally, taking the ratio of the $R_{AA}$ in different collision sizes, may give insight into the path length dependence \cite{Djordjevic:2018ita}. 

\section{Small Systems}

One of the newest frontiers of high-energy nuclear physics is the understanding on the limits of the smallest droplet of the Quark Gluon Plasma.  Relativistic hydrodynamic models reproduce collective flow observables reasonably well, however, other signatures of the Quark Gluon Plasma are not as well understood.  Even for collective flow, significant questions remain about the nature of the initial conditions \cite{Bzdak:2013zma} and the approach to hydrodynamics \cite{Strickland:2017kux,Kurkela:2018wud,Kurkela:2019set,Heinz:2019dbd}. Alternative approaches are also being explored using PYTHIA+URQMD \cite{Bierlich:2017vhg} and fluctuations derived from QCD interactions \cite{Giacalone:2019kgg}.

One fundamental question in small systems is if quarks of different flavors have sufficient time to reach thermalization.  Recently, the ALICE collaboration published a paper \cite{ALICE:2017jyt} where they found an enhancement of strangeness in small systems that could not be explained by existing models. Using effective kinetic theory, in \cite{Kurkela:2018xxd} they estimated that the minimum multiplicity to live long enough to reach thermalization would be $dN/d\eta\gtrsim 100$.  However, if one considers a core-corona model, it appears that there is still a significant contribution from the core down to $dN/d\eta\gtrsim 10$ \cite{Kanakubo:2019ogh,Kanakubo:2018vkl}.

It was originally thought that the chiral magnetic effect (CME) should only appear in large systems, however, in \cite{Khachatryan:2016got} it was found to have a significant signal in pPb collisions. Interestingly, enough it appears that when one fully incorporates electromagnetic fields in PHSD that one can obtain a splitting of charge in pPb \cite{Oliva:2019kin}. Further developments in magnetohydrodynamics have also been made \cite{Inghirami:2018ziv} that will also be relevant to future CME studies. 

While collective flow and strangeness enhancement have been measured in small systems, the suppression of hard probes and heavy flavor has not (i.e. $R_{pPb}\sim1$).  In \cite{Katz:2019qwv} an intermediate system size scan for D mesons was proposed to see the progression of $R_{AA}\rightarrow 1$ with shrinking system size and to make centrality comparisons of $v_2$, which has a non-trivial relationship with system size due to the increase in eccentricity with decreasing system size. This is an especially interest proposal considering that D mesons appear to be sensitive to out-of-equilibrium dynamics \cite{Xu:2017pna}. For the effect of quarkonium in small systems see \cite{FerreiroSQM19}.

\section{Outlook}

The study of strange and charm quarks has branched off into many new and unexpected directions that probe the fundamental theory of strongly interactions. For instance, D mesons may be used to further study far-from-equilibrium hydrodynamic behavior in small systems because of the unique information they can provide in contrast to light flow observables. For the QCD equation of state, one expects that new collaborations will spring up between heavy-ion physicists and nuclear astrophysicists who are willing to work together to better map out the QCD phase diagram at finite baryon densities. This could lead to full BSQ hydrodynamic calculations in heavy-ion collisions and the possibility of viscous fluid calculations coupled to GR in neutron star mergers. Further developments into magnetohydrodynamics and vorticity are also needed to better understand effects especially important at low beam energies. Of course, much needed context for these theoretical calculations will be provided by the Beam Energy Scan II and NICER data that are expected to appear soon.

\section*{Acknowledgments}

J.N.H. acknowledges the support of the Alfred P. Sloan Foundation and support from the US-DOE Nuclear Science Grant No. DE-SC0019175.

%
%
\bibliographystyle{h-physrev}   
\bibliography{BIG}

\end{document}